\documentclass[aps,twocolumn,superscriptaddress,showpacs,unsortedaddress]{revtex4}

\usepackage{graphicx}
\usepackage{amssymb}
\begin{document}

\title{Complex folding pathways in a simple $\beta$-hairpin}

\author{Guanghong Wei}
\affiliation{D\'epartement de physique et GCM, Universit\'e de Montr\'eal, C.P. 6128, 
succ. centre-ville, Montr\'eal (Qu\'ebec) Canada}

\author{Normand Mousseau}
\email{Normand.Mousseau@umontreal.ca}
\affiliation{D\'epartement de physique et GCM, Universit\'e de 
Montr\'eal, C.P. 6128, succ. centre-ville, Montr\'eal (Qu\'ebec) Canada}

\author{Philippe Derreumaux}
\affiliation{Information Genomique et Structurale, CNRS-UMR 1889,
 31 Chemin Joseph Aiguier, 13402 
Marseille Cedex 20, France}

\affiliation{Laboratoire de Biochimie Theorique, UPR 9080 CNRS, Institut de Biologie
Physico-Chimique, 
13 rue Pierre et Marie Curie, 75005 Paris, France}

\date{\today}

\begin{abstract}
The determination of the folding mechanisms of proteins is critical to
understand the topological change that can propagate Alzheimer and
Creutzfeld-Jakobs diseases, among others. The computational community
has paid considerable attention to this problem; however, the
associated time scale, typically on the order of milliseconds or more,
represents a formidable challenge. {\it Ab initio} protein folding
from long molecular dynamics (MD) simulations or ensemble dynamics is
not feasible with ordinary computing facilities and new techniques
must be introduced. Here we present a detailed study of the folding of
a 16-residue $\beta$-hairpin, described by a generic energy model and
using the activation-relaxation technique.  From a total of 90
trajectories at 300 K, three folding pathways emerge. All involve a
simultaneous optimization of the complete hydrophobic and hydrogen
bonding interactions. The first two follow closely those observed by
previous theoretical studies. The third pathway, never observed by
previous all-atom folding, unfolding and equilibrium simulations, can
be described as a reptation move of one strand of the $\beta$-sheet
with respect to the other. This reptation move indicates that
non-native interactions can play a dominant role in the folding of
secondary structures. These results point to a more complex folding
picture than expected for a simple $\beta$-hairpin.


\end{abstract}

\maketitle


\section*{\bf INTRODUCTION}
As one of the smallest building blocks of proteins, the $\beta$-hairpin 
and particularly the second $\beta$-hairpin of the domain B1 of protein 
G (referred to as $\beta$-hairpin2) has been the subject of many 
theoretical and experimental folding studies. This peptide adopts hairpin
structures in solution but overall its flexibility 
precludes the determination of a high-resolution NMR solution 
structure.~\cite{NMR95} Fluorescence experiments show that this 
$\beta$-hairpin folds in isolation with a time constant of 6 microseconds 
and its folding kinetics is described by the two-state
model.~\cite{EATON} Because these data do not provide details of the
transition and such a time scale cannot be covered by
hundreds of long molecular dynamics (MD)-trajectories at 300 K in explicit 
solvent,~\cite{DU98} alternative methods have been used in order to
characterize the thermodynamics and folding kinetics of the 
$\beta$-hairpin2.

Two folding mechanisms have already  been proposed. 
The first mechanism, suggested by statistical mechanical 
models~\cite{MUN98} and lattice Monte Carlo (MC) simulations,~\cite{KI99} 
is that folding starts at the turn and propagates towards the tail by 
hydrogen bonding interactions, the hydrophobic cluster forming at the 
end of folding. One variant of this mechanism suggested by Langevin 
dynamics of an off-lattice model is that the formation of the hydrophobic 
cluster is followed by zipping of hydrogen bonds (H-bonds), predominantly
starting from those near the turn.~\cite{THI00}
Another variant suggested by all-atom MD simulations is that the
$\beta$-hairpin
folds beginning at the turn, followed by hydrophobic collapse and then
H-bond formation.~\cite{TSAI02}

The second mechanism proposed is that the N- and C-termini first approach 
each other to form a loop, and the structure propagates from there.
This mechanism is apparently independent of all-atom force field details, 
since it has been recognized by ensemble dynamics at 300 K using 
implicit solvent,~\cite{PD01} replica exchange method combining MD
 trajectories
with a temperature exchange MC process using SPC model 
solvent,~\cite{BS01} minimalist Go-folding discontinuous MD 
simulations,~\cite{ZHOU02} unfolding simulations,~\cite{PD99,LEE01} and 
multicanonical MC simulations with an implicit solvent model.~\cite{MK99}

Along with the difference in folding dynamics within each scenario, three 
questions are yet to be resolved. The first question is when the native
H-bond network and hydrophobic core form: (i) the hydrophobic core is being
 formed
first, and the H-bonds appear,~\cite{GS01,PD99,LEE01,MK99,MA00} (ii)
 the H-bonds
form first and then the hydrophobic core,~\cite{EATON} or (iii) the final
hydrophobic core and H-bonds form simultaneously.~\cite{ZHOU02,PD01,THI00}
The second question is whether helical 
structures exist during folding process. Berne et al., by using the
OPLSAA force field, did not find evidence of significant helical structures
 in their simulations at all temperatures studied,~\cite{BS01} while
 Garcia and 
Sanbonmatsu found a helical content of 15\% at low temperatures,~\cite{GS01} 
 Pande et al. detected short-lived semi-helical intermediates at room
 temperature,~\cite{PD01} and Irb\"ack found a low population of $\alpha$-helix 
structure at 273 K.~\cite{IRB03}  
The third question is whether the previously used methods, which fail to 
detect both pathways, may miss other major folding pathways, as has been 
discussed recently~\cite{AF02} for ensemble dynamics which uses a large 
number of short MD simulations of only tens of nanoseconds and a 
supercluster of thousands of computer processors.
 
To address these issues, we simulate the folding mechanisms of hairpin2  
using a previously described model --- the Optimized Potential for Efficient 
peptide-structure Prediction (OPEP), and sampling technique --- the 
Activation-Relaxation Technique (ART).~\cite{BM96, BM98} 
OPEP  can be used to 
simulate any amino acid sequence and works for proteins that do 
and do not form ordered structures in solution. Ordered structures include 
three-helix bundles and three-stranded anti-parallel $\beta$-sheet structures, 
among others.~\cite{FOR01} 
For its part, ART generates trajectories on the configurational energy
 landscape, 
identifying a series of energy minima separated by first-order saddle points.
  The 
efficiency of ART is not affected by 
the height of the activation energy barriers or the complexity of the atomic 
rearrangements and thus samples very efficiently the rugged-energy 
landscapes of small proteins.
The OPEP-ART approach has been applied recently to study the 
folding of three sequences adopting an $\alpha$-helix, a three
 stranded antiparallel 
$\beta$-sheet and a $\beta$-hairpin helix in solution.~\cite{WMD02} 
In this work, 82 folding simulations at 300 K 
start from a fully extended conformation ($\phi$= $-$180$^{\circ}$, 
$\psi$= 180$^{\circ}$) using different random-number seeds: 52 use the 
standard OPEP force field (16 reaching the folded state), 20 use a modified 
set of OPEP parameters (6 reaching the folded state), and 10 use a biased 
Go-like potential (10 reaching the folded state). To determine the effect
 of the 
starting structure, we also launched 8 independent runs
 (4 reaching the folded 
state) at 300 K starting from a semi-helical conformation using the standard 
OPEP force field. 

From a total of 90 trajectories at room temperature, 36 found 
the native state providing a detailed picture of the folding mechanism.
Although all these folding trajectories involve a simultaneous optimization of 
the complete hydrophobic and hydrogen bonding interactions, the 36 folding
 runs can be described by 3 mechanisms: two of them follow closely
 those observed 
by previous theoretical and computational studies, but the third one
 represents a new 
folding mechanism for proteins. This mechanism can be described
as a reptation move of one strand of the $\beta$-sheet with respect to
the other. These three mechanisms  offer  
a complete picture of the $\beta$-hairpin folding, independently of the 
exact amino acid composition, and help reconcile conflicting theoretical 
data on the hairpin2 of protein G~\cite{EATON,MK99,THI00,ZHOU02} or 
between various hairpins, e.g., the first hairpin of tendamistat~\cite{BON00}
and a 11-residue model peptide.~\cite{WJ99} The existence of these
three competing mechanisms was presented recently in a short
communication;~\cite{WMD03} here, we offer a detailed description of
the folding mechanisms in this simple $\beta$ hairpin. Furthermore, we
 present
the results of new simulations using either G\=o-like potential or
 semi-helical
starting conformations.

\section*{\bf METHODS}

We have simulated the folding of the C-terminal $\beta$-hairpin from
protein G (residues 41-56).  The sequence of the peptide is
GEWTYDDATKTFTVTE. The energy surface was modeled using the OPEP model
and the dynamics was obtained by the activation-relaxation technique.

\subsection*{\bf Activation-Relaxation Technique}

ART is a generic method to explore the landscape of continuous energy
functions through a series of activated steps. The algorithm has
evolved considerably over the years and here we apply its most recent
version, ART nouveau,~\cite{MM00,MDB01} which uses a recursion method,
the Lancz\'os algorithm,~\cite{Lan88} to extract the direction of lowest
curvature of the landscape leading to a first-order saddle point.
Such an approach provides an efficient way to extract a limited
spectrum of eigenvectors and eigenvalues without requiring the
evaluation and diagonalization of the full Hessian matrix. A similar 
approach was also introduced in Ref.~\onlinecite{Wales99}.
An ART event is defined directly in the space of configurations,
which allows for moves of any complexity, and consists of four steps:
 
\begin{enumerate}
\item Starting from a local minimum, a configuration is first pushed
outside the harmonic well until a negative eigenvalue appears in the
Hessian matrix.

\item The configuration is then pushed along the eigenvector
associated with the negative eigenvalue until the total force is
close to zero, indicating a saddle point. The first two steps constitute
the activation phase.

\item The configuration is pushed slightly over the saddle point and
is relaxed to a new local minimum, using standard minimization
technique.

\item Finally, the new configuration is accepted/rejected using the 
Metropolis criterion at the desired temperature. 
In each of the simulations at hand, this four-step procedure was
repeated 4000 times, taking less than 18 processor-hours on an IBM Power-4
machine
\end{enumerate}

As discussed in our previous work,~\cite{WMD02} the temperature in ART
is not a real temperature since ART samples the conformational space from 
one minimum to another minimum. However, ART generates 
well-controlled trajectories (more than 83\% of events relax back to within 
0.4~\AA\ from their initial minima starting at the saddle points).~\cite{WMD02}
A detailed description of the algorithm and implementation of 
ART can be found in earlier publications.~\cite{BM96,BM98,WMD02}

\subsection*{\bf Energy Model}

We use a coarse-grained off-lattice model where each amino acid is
represented by its N, H, C$\alpha$, C, O and one bead for its side
chain.  The exact OPEP energy function, which includes solvent effects
implicitly, was obtained by maximizing the energy of the native fold
and an ensemble of non-native states for six training peptides with
10-38 residues. In this work, the side chain propensities of the 20
amino acids for $\alpha_R$ helix, $\beta$-strand and $\alpha_L$ helix
~\cite{DP99,DP00} are neglected.  The total energy is thus expressed
by:
\begin{eqnarray}
E &=& w_{L} E_L + w_{H} E_{HB1} + w_{HH} E_{HB2} + w_{SC,SC} E_{SC,SC}
 \nonumber \\
& & + w_{SC,M} E_{SC,M} + w_{M,M} E_{M,M}
\end{eqnarray}

The interaction potentiel OPEP is a function of the weights $w$'s of
the following interactions:

\noindent
(i) quadratic terms to maintain stereochemistry: bond lengths and bond
angles for all particles and improper dihedral angles for the side
chains and the peptide bonds $E_L$,

\noindent
 (ii) and excluded-volume potential of the main chain interactions $
  E_{M,M}$ and of side chain--main chain interactions $ E_{SC,M}$,

\noindent
 (iii) pairwise contact 6-12 interactions between the side chains
considering all 20 amino acid types $E_{SC,SC}$,

\noindent
 (iv) backbone two-body $E_{HB1}$ and four-body $E_{HB2}$ hydrogen
bonding interactions. All nonbonded interactions are included (no
truncation).

The two-body energy of one H-bond between residues $i$ and $j$ is
defined by
\begin{eqnarray}
E_{HB1} &=& \varepsilon_{hb} \sum_{ij}\mu(r_{ij})\nu(\alpha_{ij})
\end {eqnarray}
where
\begin{eqnarray}
\mu(r_{ij}) &=& 
5(\frac{\sigma}{r_{ij}})^{12}-6(\frac{\sigma}{r_{ij}})^{10}
\end {eqnarray}

\begin{eqnarray}
\nu(\alpha_{ij}) &=& \left\{\begin{array}{r@{\quad \ \ \ 
\quad}}\cos^{2} \alpha_{ij} \ \ \ \ \ \ \ \  \alpha_{ij} > 90^{\circ}
\\ 0 \ \ \ \ \ \ \ \ \ \ \ \  otherwise
\end{array} \right.
\end{eqnarray}
where $r_{ij}$ is the O..H distance between the carbonyl oxygen and
amide hydrogen and $\alpha_{ij}$ the NHO angle.

The cooperative energy between two neighbored H-bonds $ij$ and
$kl$ is defined by
\begin{eqnarray}
E_{HB2} &=& \varepsilon_{2hb} \exp(-(r_{ij}-\sigma)^2/2)
\exp(-(r_{kl}-\sigma)^2/2)
 \nonumber \\
& &\Delta(ijkl)  
\end {eqnarray}
where $\Delta(ijkl)$ = 1 if (k,l) = (i+1, j+1) or (i+2, j-2) or
(i+2, j+2), otherwise $\Delta(ijkl)$ = 0. This corresponds to
the pattern of H-bonds in $\alpha$-helices, anti-parallel and
parallel beta-sheets, respectively.

In this work, unless specified we use $\sigma$ = 1.8~\AA,
$\varepsilon_{hb}$ = 1.0 kcal/mol if j=i+4 (helix), otherwise = 1.5
kcal/mol, and $\varepsilon_{2hb}$ = $-$2.0. 
The other parameters can be found in Ref.~\onlinecite{FOR01}.

\subsection*{\bf Trajectory Analysis}

Our native structure contains six main chain H-bonds excluding the one
at the turn since it rarely forms because of geometrical
constraints. Following Karplus ~\cite{MK99} and Garcia,~\cite{GS01}
they are numbered from the tail to the turn 42:HN-55:O (H1), 55:HN-42:O (H2),
 44:HN-53:O (H3), 53:HN-44:O (H4), 46:HN-51:O (H5),
51:HN-46:O (H6). The expression $i$:HN-$j$:O denotes the atoms
involving a H-bond between residues $i$ and $j$.

To characterize a conformation, we use the number of native H-bonds, 
the radius of gyration of the hydrophobic core
($Rg_{core}$), and the $C_\alpha$-rmsd of residues 41-56 from the 2GB1
structure.\cite{GF91} $Rg_{core}$ is calculated using the side chains
of the four residues W43, Y45, F52, and V54. A H-bond is defined if it
satisfies DSSP conditions:~\cite{KS83} the distance between the
carbonyl oxygen and amide hydrogen (O..H) is less than 2.4~\AA\, and
the NHO angle is greater than 145$^\circ$.

The regions of conformational space that have been sampled by all
simulations are clustered as follows. The C$_{\alpha}$-rmsd is calculated 
for each pair of structures in each simulation. The number of neighbors 
is then computed for each structure using a C$_{\alpha}$-rmsd cutoff of 
1.5 \AA. The conformation with the highest number of neighbors is 
considered as the center of the first cluster. All the neighbors of this 
conformation are removed from the ensemble of conformations. The center 
of second cluster is then determined in the same way as for the first cluster, 
and this procedure is repeated until each structure is assigned to a cluster.
Then we have a list of central structures of clusters for every folding simulation.
Once all the trajectories are clustered, we can cluster all the central structures 
in two different folding trajectories, in order to identify the common clusters 
between the simulations.

\section*{\bf RESULTS}
\subsection*{\bf Native vs. Non-native Hairpin Structures}

Cluster analysis of all ART-generated trajectories shows that the
lowest-energy conformation ($\it E$ = $-$ 33 kcal/mol) deviates by
less than 1 \AA~ $C_\alpha$ rms from the hairpin structure within
protein G (PDB code 2GB1~\cite{GF91}). For the purpose of our
simulation, we define a structure as being fully folded -- or native
-- if it satisfies the following four criteria: (i) the six native
H-bonds are formed (see Methods); (ii) the
backbone dihedral angles ($\phi$, $\psi$) have standard $\beta$-sheet
values (around ($-$90$^{\circ}$, 150$^{\circ}$)); (iii) the
 hydrophobic core is well packed
(core radius of gyration is around 4.3 \AA) and (iv) the all-residue
$C_\alpha$ rmsd from 2GB1 is less than 2.5 \AA.  These conditions for
nativeness are thus more stringent than those of in previous folding
simulations.~\cite{PD01, ZHOU02} For Zagrovic and
collaborators,~\cite{PD01} the H-bonds connecting the two
strands are not required to be native and hairpin with asymmetric
strands are considered as native structures. Similarly, for Zhou {\it
  et al},~\cite{ZHOU02} the condition for nativeness is that all-heavy
atom rmsd from the global minimum structure is less than 2.5 \AA. The
rmsd between their modeled structure and the experimental 2GB1 is not
given.

\begin{figure*}[ht!]
\includegraphics[width=14cm]{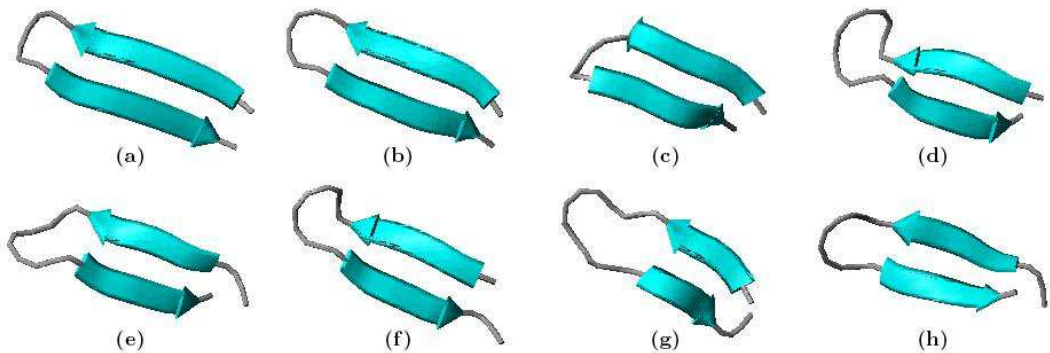}
\vspace{0.0cm}
\caption{
The sampled hairpin structures with different main chain 
H-bonds and 2GB1 structure. (a) 2GB1 structure; (b) Hairpin structure 
with 6 native H-bonds (H1-H6), rmsd =1.09 (0.88)~\AA, E = -33 kcal/mol; 
(c) hairpin structure with 5  key H-bonds (H1-H4, 52:HN-45:O),
rmsd = 2.87 (2.15) \AA, E = -29 kcal/mol; (d) hairpin structure
 with 4 key hydrogen 
bonds (43:HN-54:O, 45:HN-52:O, 52:HN-45:O, 54:HN-43:O), rmsd = 2.62 (2.39) 
\AA, E = -28 kcal/mol; (e) hairpin structure with 4 key H-bonds 
(44:HN-54:O, 46:HN-52:O, 52:HN-46:O, 54:HN-44:O), rmsd = 2.08 (1.87)~\AA,
E = -30 kcal/mol; (f) hairpin structure with 5 key H-bonds (43:HN-53:O, 
45:HN-51:O, 51:HN-45:O, 53:HN-43:O, 45:HN-41:O), rmsd = 2.42 (2.18)~\AA, 
E = -29 kcal/mol; (g) hairpin structure with 4 key H-bonds (42:HN-54:O, 
44:HN-52:O, 52:HN-44:O, 54:HN-42:O), rmsd = 2.40 (1.98)~\AA, E = -28 kcal/mol; 
(h) hairpin structure with 4 key H-bonds (43:HN-55:O, 45:HN-53:O, 
53:HN-45:O, 55:HN-43:O), rmsd = 3.56 (2.96)~\AA, E = -28 kcal/mol. The
rmsd value in parentheses is the C$_\alpha$-rmsd from 2GB1 for residues 
43-54. According to our definition of folded state, only hairpin (b) is
 folded state. 
}
\label{fig:hairpins}
\end{figure*}

Figure~\ref{fig:hairpins} (produced by using the MOLMOL software \cite{MOL96}) 
shows the 2GB1 structure (a), our native
structure (b) and six non-native hairpin structures (c-h) sampled by
ART.  Figure~\ref{fig:hairpins}(e) and ~\ref{fig:hairpins}(h) show two
hairpins in which the turn (residues 7-11) is shifted toward C
terminus. Figure~\ref{fig:hairpins}(f) and \ref{fig:hairpins}(g) show
two non-native hairpins where the turn (residues 6-10) is shifted
toward the N terminus. In this study, only hairpin (b), of lowest
energy, is the native state, while, according to the definition of
folded state in Ref.~\onlinecite{PD01}, hairpin structures (c)-(h) are
also folded states. Our hairpin structures (c) and (d) are very
similar to the folded state in Series 17 (two key H-bonds are
HB53-44 and HB52-45) and Series 2 (two key H-bonds are HB45-52
and HB43-54) of Ref.~\onlinecite{PD01}, respectively. Both of them have
symmetric strands, but different H-bond network
pattern. Moreover, from the key H-bonds in their eight
independent folding trajectories, it seems that the hairpin
structures in Series 1, 7, 9, 11 are asymmetric about the
$\beta$-turn.

\begin{figure*}[ht!]
\includegraphics[width=14cm]{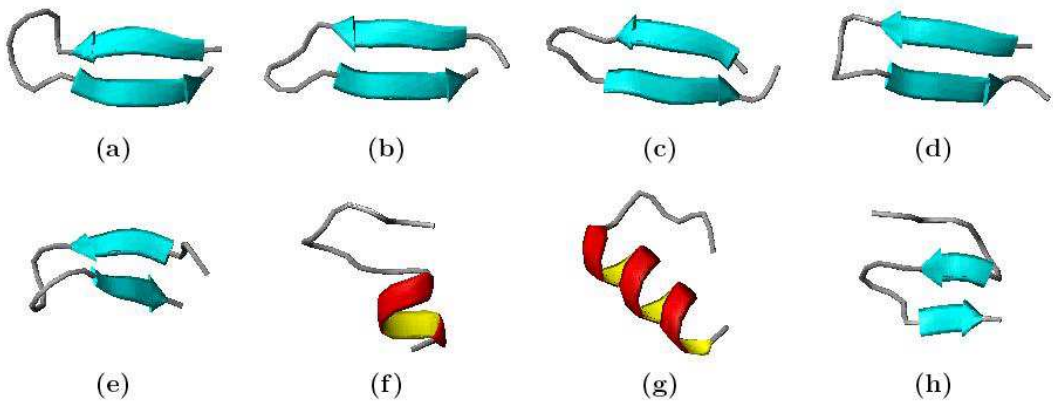}
\vspace{0.0cm}
\caption{
The sampled metastable states using the standard OPEP potential,
starting from a fully extended state. (a) hairpin structure with
 4 key H-bonds 
 (43:HN-54:O, 45:HN-52:O,52:HN-45:O, 54:HN-43:O), rmsd = 2.62~\AA, 
E = -28 kcal/mol; (b) hairpin structure with 4 key H-bonds (44:HN-54:O, 
46:HN-52:O, 52:HN-46:O, 54:HN-44:O), rmsd = 2.08~\AA, E = -30 kcal/mol; 
(c) hairpin structure with 4 key H-bonds (44:HN-54:O, 46:HN-52:O, 
52:HN-46:O, 54:HN-44:O), rmsd = 2.88~\AA, E = -27 kcal/mol; (d) hairpin 
structure with 5 key H-bonds (43:HN-53:O, 45:HN-51:O, 51:HN-45:O, 
53:HN-43:O, 45:HN-41:O), rmsd = 2.6~\AA, E = -29 kcal/mol; (e) hairpin 
structure with 4 key H-bonds (44:HN-55:O, 46:HN-53:O, 53:HN-46:O, 
55:HN-44:O), rmsd = 3.34~\AA, E = -23 kcal/mol; (f) a $\alpha$-$\beta$
structure with short helix appearing in the C terminal, rmsd = 7.39~\AA, 
E = -22 kcal/mol; (g) an $\alpha$-helix structure involving residues from 5
 to 15, rmsd = 5.89~\AA, E = -23 kcal/mol; (h) a three-stranded beta-sheet structure 
with 4 H-bonds (42:HN-49:O, 49:HN-42:O, 48:HN-54:O, 54:HN-48:O), 
rmsd = 7.81~\AA, E = -27 kcal/mol. Two metastable state (c) and (d) converge 
to the native state within 5000 additional trial events.
}
\label{fig:non-native}
\end{figure*}

\subsection*{Analysis of Unfolded Trajectories}

From  a total of 90 runs, 36 trajectories reach the native state. 
The other 54 fail to locate the native
structure within 4000 ART-events. These runs lead either to non-native
hairpin conformations as discussed previously or to other metastable
conformations of various secondary compositions, e.g. $\alpha$-helix
with coil , three-stranded antiparallel $\beta$-sheet and short
$\alpha$-helix with $\beta$-like structures. 

\begin{figure}[ht!]
~\\ \vspace{-6.0cm}\includegraphics[width=8.8cm]{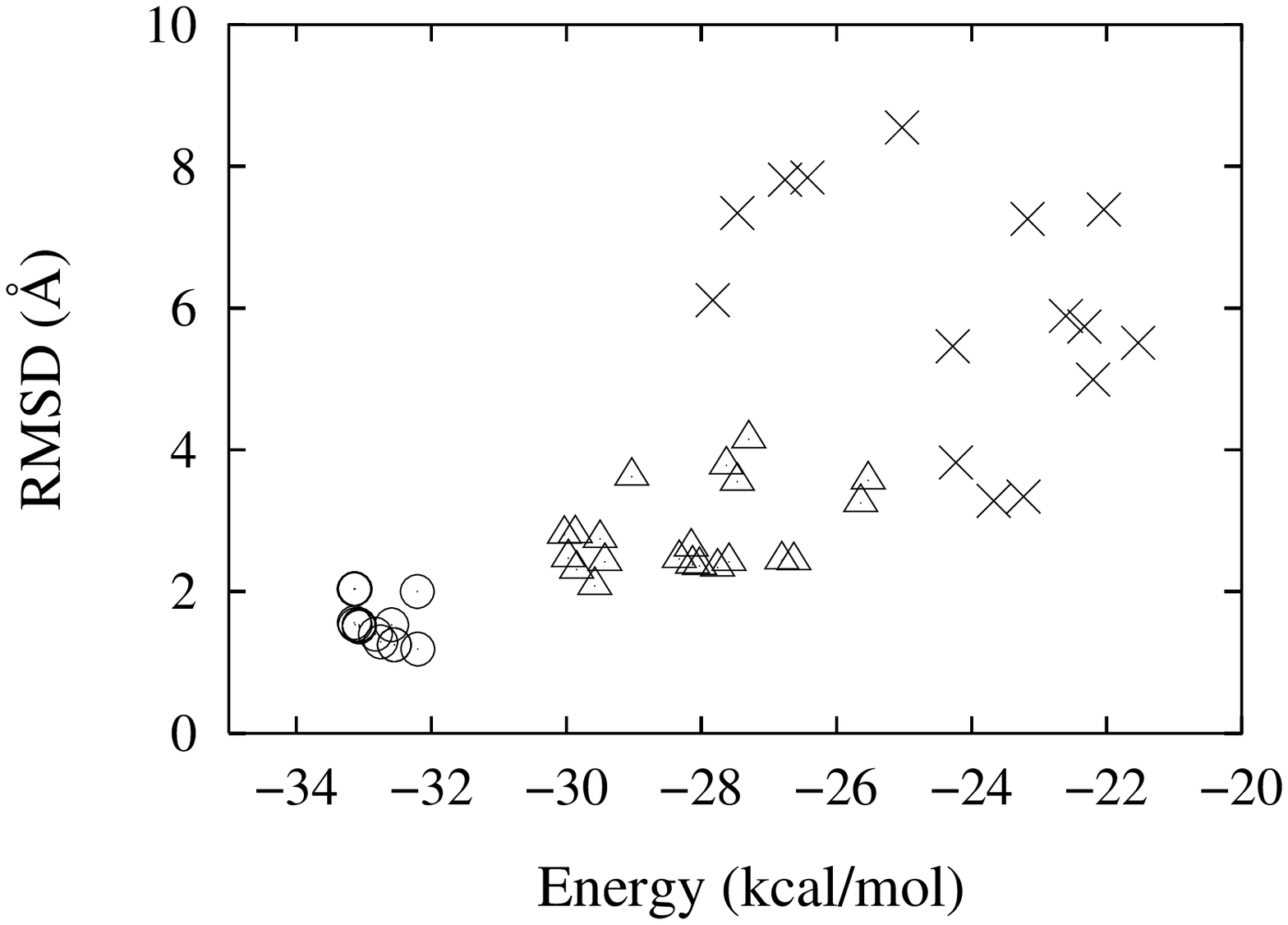}
\vspace{0.0cm}
\caption{
Rmsd as a function of energy of the lowest-energy structures 
generated in the 52 simulations (16 folded and 36 unfolded) using the 
standard OPEP potential, starting from a fully extended state. Circles,
triangles, and crosses are  for native $\beta$-hairpin, asymmetric $\beta$-hairpin with 
different H-bond pattern, and other different structures, respectively. 
}
\label{fig:rmsd-E}
\end{figure}

Figure~\ref{fig:non-native} shows the structural features
of  the 8 lowest energy metastable structures sampled. The 54 metastable
states lie between $-$21 and $-$30 vs. $-$33 kcal/mol for the native
state. 
It is important to note that these structures do not represent dead
ends for the simulation; continuing the simulation at 300~K
for two arbitrarily chosen metastable states (c)
and (d), we find that it is possible to reach the fully formed hairpin
structure within 5000 additional trial events.
In Fig.~\ref{fig:rmsd-E}, we plot the energy vs. rmsd for the
lowest-energy structures in all the 52 simulations (16 folded and 36
unfolded) using the standard OPEP and starting from the fully extended
state. We see that all folded structures appear in a dense region
around $-$33 kcal/mol and below an rmsd of 2.0~\AA\ and are well
separated from the non-folded ones; clearly, our potential can
discriminates folded states from metastable states.  By visual
inspection of the 36 metastable states and comparison with 
Fig.~\ref{fig:rmsd-E}, we see (i) that the conformations
with $E$ between $-$25 and $-$30 kcal/mol and rmsd between
2.0 and 4.1~\AA\ adopt asymmetric $\beta$-sheet structures with
different H-bond networks, (ii) that the conformations with $E$ 
between $-$25 and $-$28 kcal/mol and rmsd between 6.0 and
8.5~\AA\ show three-stranded antiparallel $\beta$-sheet structures,
 (iii) and that the other conformations with $E$ between
$-$21 and $-$25 kcal/mol and rmsd between 3.0 and 8.0~\AA\ adopt 
$\alpha$-helix with coil, short $\beta$-hairpin with coil, or short
$\alpha$-helix with $\beta$-like structures. 
As the rmsd of some non-native hairpin structures can be as small as
2.1 \AA\ (see Fig.~\ref{fig:non-native}(b)), it is clear that this
 sole criterion 
is not sufficient to differentiate between native and non-native hairpin 
structures.

\subsection*{Analysis of Folded Trajectories}

As reported in Ref.~\onlinecite{WMD03}, the 16 folding simulations using the
standard OPEP potential which start from a fully extended state can be
classified into 3 mechanisms : I (4 runs), II (7 runs) and III (5
runs).

\begin{figure*}[ht!]

\vspace{-0.6cm}\includegraphics[width=13cm]{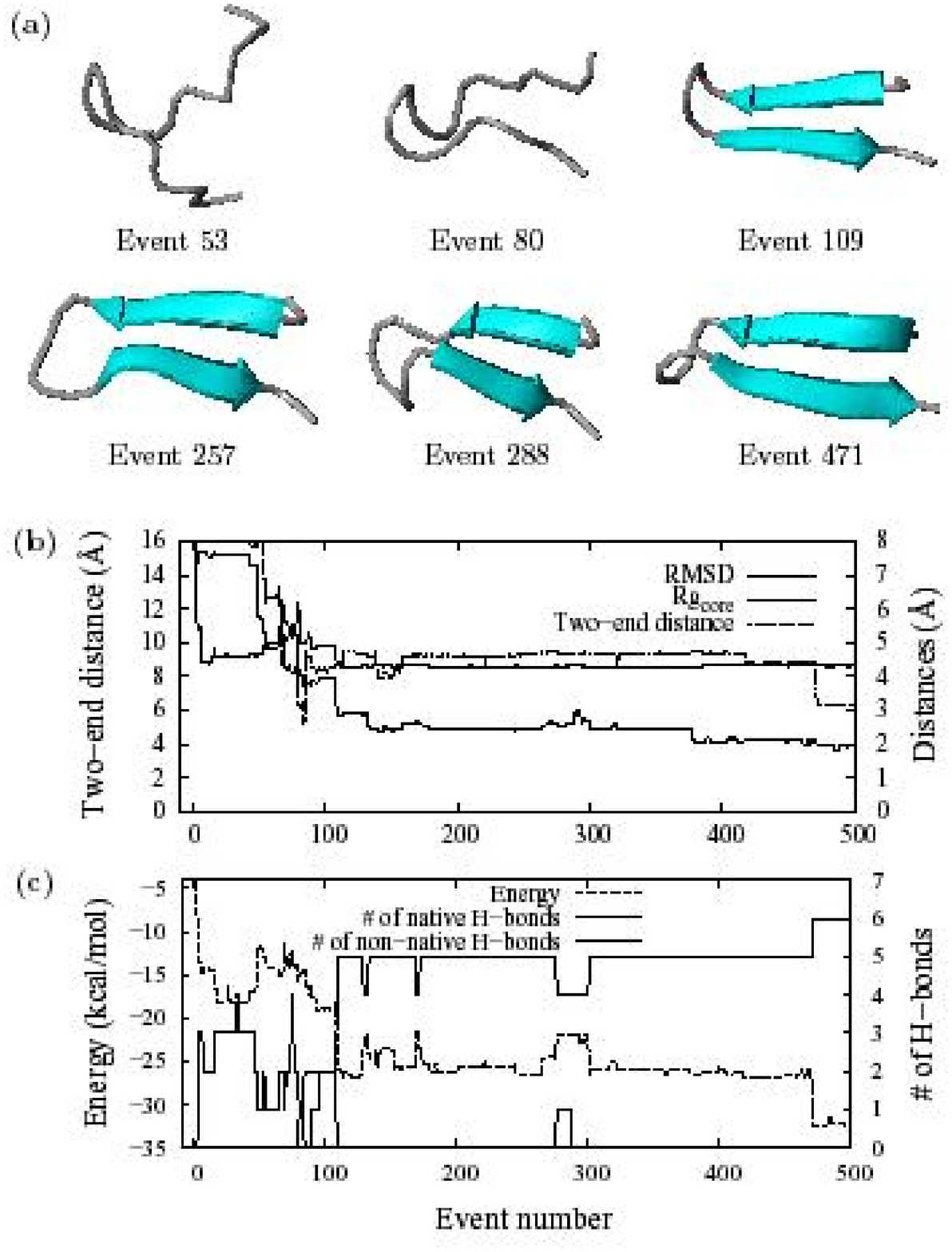}

\vspace{0.0cm}
\caption{
Detailed analysis of a representative folding trajectory following Mechanism I,
simulated at 300 K, starting from a fully extended state. (a) Six snapshots.  
Only accepted events are shown. (b) C$_\alpha$-rmsd from 2GB1 structure of 
the hairpin, radius of gyration of the hydrophobic core $Rg_{core}$, and the 
two-end distance as a function of accepted event number. (c) Total energy, number 
of all the native and non-native H-bonds in each sampled conformation as a function 
of accepted event number.
}
\label{fig:rmsd-extend-300k-s4}
\end{figure*}
\begin{figure}[ht!]
\vspace{0.0cm}\includegraphics[width=8.6cm]{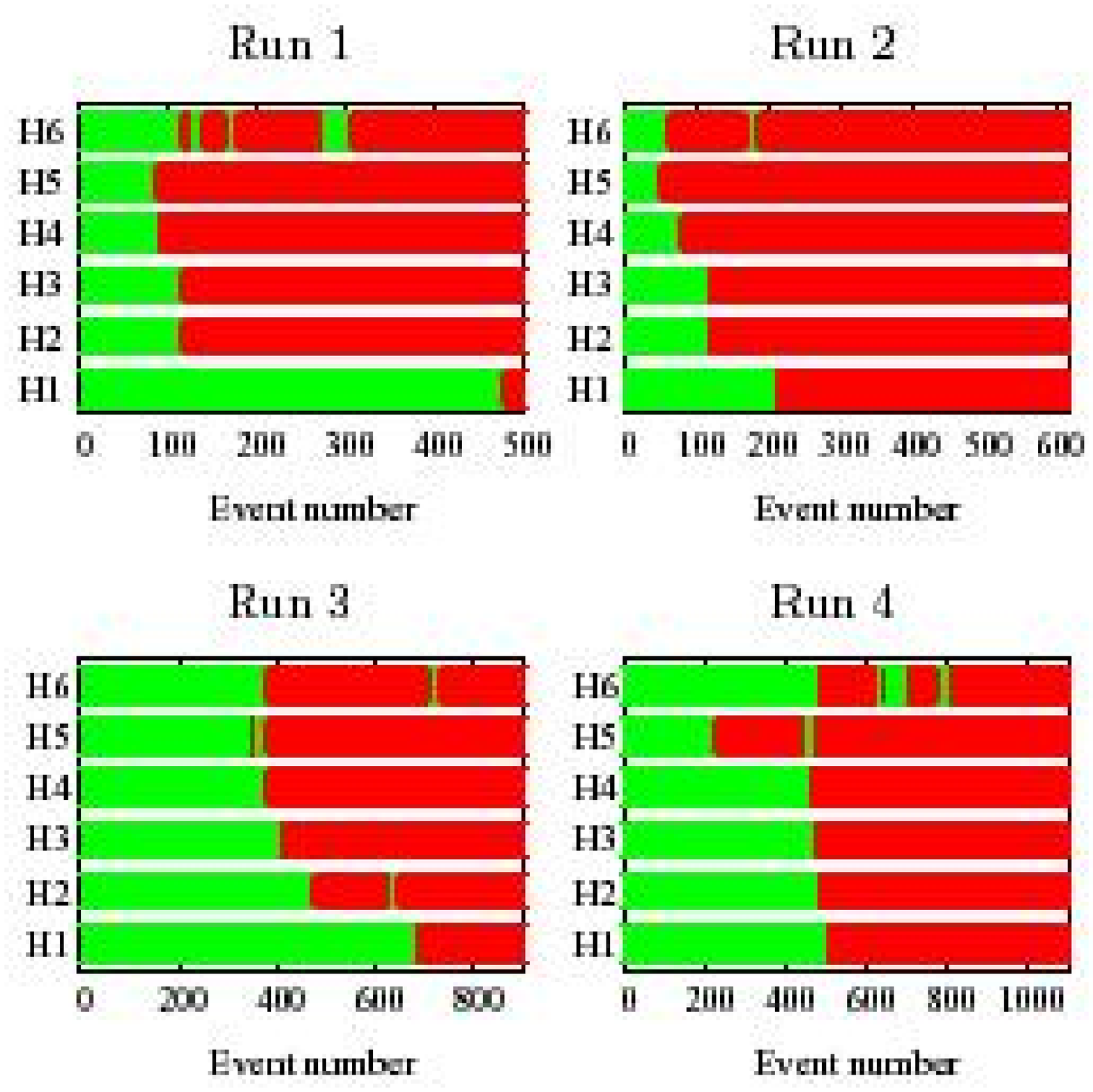}
\caption{
Status of the six native H-bonds as a function of accepted event number 
in the four runs following Mechanism I. Green: not formed, red: formed. Run 1 is the 
simulation described in Fig. 4.
}
\label{fig:hb-mechanism-I}
\end{figure}

A detailed analysis of mechanism I, seen in four folding trajectories,
is given in Figure~\ref{fig:rmsd-extend-300k-s4}. This mechanism is
similar to that described in
Refs.~\onlinecite{KI99, THI00, WJ99}. Starting from a fully extended
state, the peptide first collapses into a compact state with the turn
placed in the right section of the chain (residues 7-10) (Events 53);
this step is characterized by the formation of a partially packed
hydrophobic core (radius of gyration of the hydrophobic core, 
$Rg_{core}$, is 4.8 \AA, see Fig.~\ref{fig:rmsd-extend-300k-s4}(b)) 
and the appearance of several
non-native H-bonds (Fig.~\ref{fig:rmsd-extend-300k-s4}(c)).  The
following steps serve to stabilize its hydrophobic core. At event 80,
a native H-bond near the turn (H5) forms, followed rapidly by
the formation of H4; 29 events later (event 109), the peptide
reorganizes its hydrophobic core to a well packed state ($Rg_{core}$ 
reaches its final value 4.3~\AA). The
reorganization of the hydrophobic core allows the formation of new
native main chain H-bonds (H3, H2, H6). At that point, the
two-end distance fluctuates around 9~\AA\
(Fig.~\ref{fig:rmsd-extend-300k-s4}(b)). In spite of these new
native H-bonds, the flexibility of the loop remains large and H6, an
H-bond near the loop, breaks and reforms many times between
events 128 and 300 (see Run 1 in Fig.~\ref{fig:hb-mechanism-I}). 
Finally, the  H-bond H1, near the end of the
peptide, forms at last (event 471) leading to the native state: the
core radius of gyration remains its final value 4.3 ~\AA, the rmsd
from 2GB1 structure drops to 2.0~\AA\
(Fig.~\ref{fig:rmsd-extend-300k-s4}(b)), the total number of native
H-bonds is six, and the total energy is $-$32 kcal/mol
(Fig.~\ref{fig:rmsd-extend-300k-s4}(c)). The time formation of
the six native H-bonds H1-H6 can be seen in
Fig.~\ref{fig:hb-mechanism-I} for the four folding runs.
The number of native and non-native H-bonds in
each accepted conformation as a function of event number is given in
Fig.~\ref{fig:rmsd-extend-300k-s4}(c). We clearly see that 
the folding process is a competition between native and non-native 
hydrogen bonding interactions, and that in Mechanism I the two ends
of this $\beta$-hairpin gradually come near to form H1 last, and the
hydrophobic core becomes well packed rapidly.

\begin{figure*}[ht!]
\vspace{-0.6cm}\includegraphics[width=13cm]{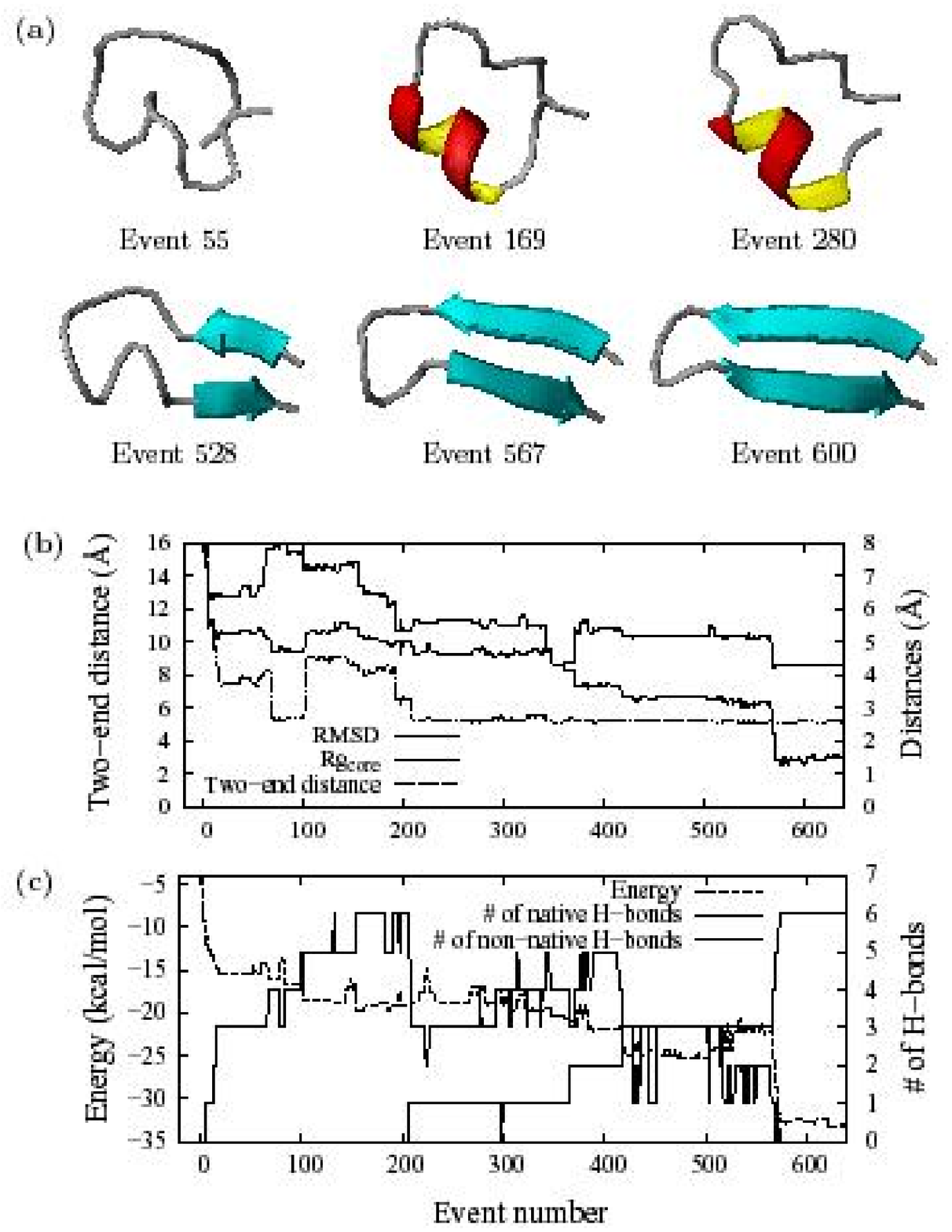}
\vspace{0.0cm}
\caption{
Detailed analysis of a representative folding trajectory following Mechanism II,
simulated at 300 K, starting from a fully extended state. (a)  Six snapshots. 
(b) C$_\alpha$-rmsd from 2GB1 structure of 
the hairpin, radius of gyration of the hydrophobic core $Rg_{core}$, and the 
two-end distance as a function of accepted event number. (c) Total energy,
 number of all the native 
and non-native H-bonds in each sampled conformation as a function of
accepted event number. 
}
\label{fig:rmsd-extend-300k-s5}
\end{figure*}

\begin{figure}[ht!]
\vspace{0.cm}\includegraphics[width=8.6cm]{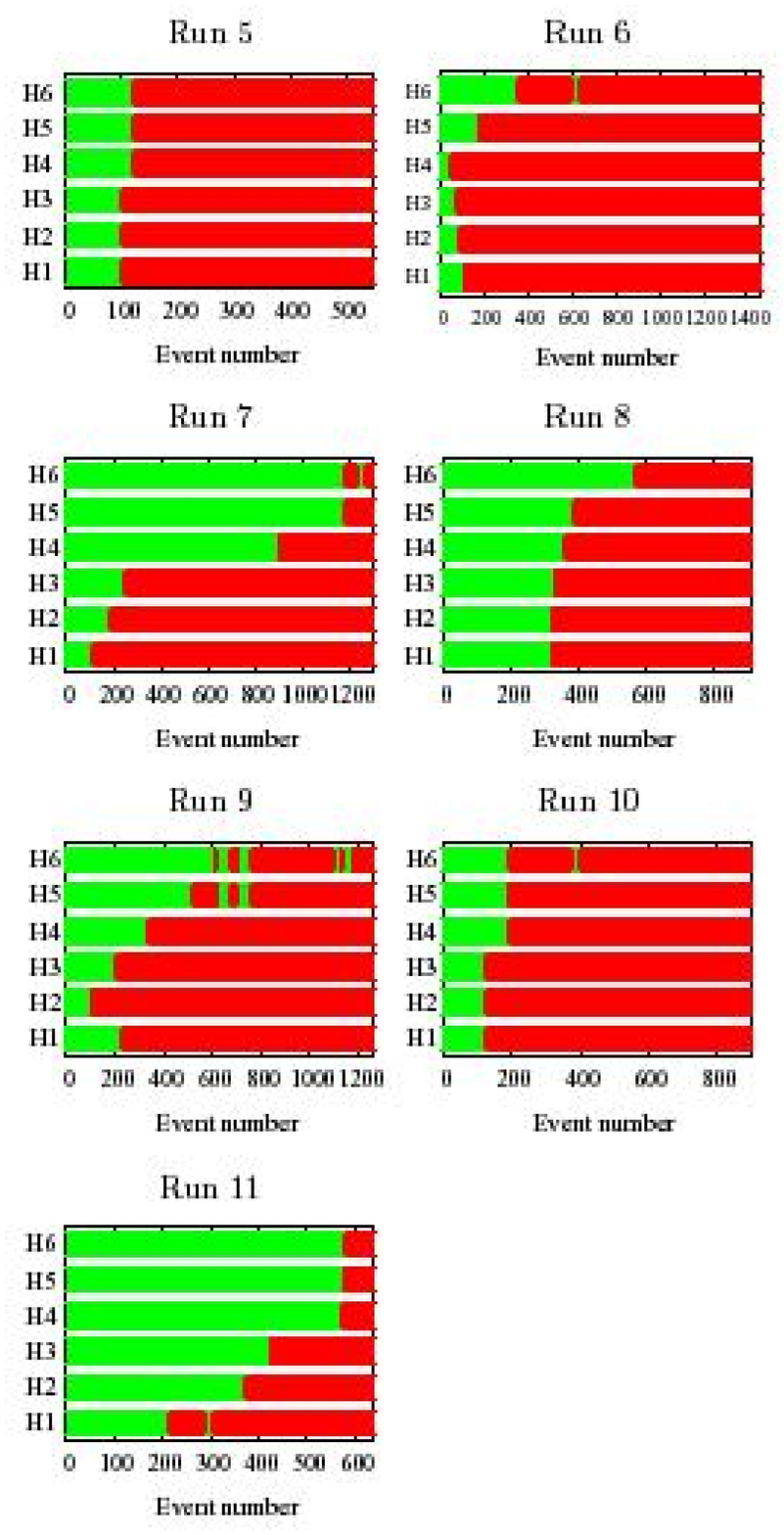}
\vspace{0.0cm}
\caption{
Status of the six native H-bonds as a function of accepted event number in 
the seven runs following Mechanism II. Green: not formed, red: formed. Run 
11 is the simulation described in Fig. 6.
}
\label{fig:hb-mechanism-II}
\end{figure}

Mechanism II, seen in seven folding trajectories, was observed in
simulations of hairpin2~\cite{PD01,BS01,MK99} and of a 10-residue
model peptide.~\cite{NK02} Figure~\ref{fig:rmsd-extend-300k-s5} shows
the major folding steps associated with this mechanism. Within a few
events, the hydrophobic interaction between the four residues W43,
Y45, F52, and V54 induces the formation of a partial hydrophobic core
($Rg_{core}$ is 6.4~\AA) resulting in a globular
state with three non-native H-bonds (event 55). At the same time, the
N- and the C-termini approach each other to form a large loop.
Because of the competition between the hydrophobic, native, and
non-native hydrogen bonding forces, the peptide rearranges its
hydrophobic core by breaking and reforming some non-native
interactions ($Rg_{core}$ increases to 8~\AA). At
event 191, the hydrophobic core drops to 5.5~\AA\ and, after 16 more
events, the H-bond H1 (event 207) near the end forms. This
reorganization of hydrophobic core causes the second (H2 at event 366)
and third (H3 at event 417) H-bonds to form.  After a 150-event optimization 
between hydrophobic and hydrogen bonding interactions, the radius of 
gyration of the hydrophobic core reaches its final value (4.3~\AA\ at event 
567), and the fourth H-bond H4 appears. In the following
several events, H5 (at event 570) and H6 (at event 573) form; the
peptide satisfies the native conditions (see Fig.~\ref{fig:rmsd-extend-300k-s5}). 
Fig.~\ref{fig:hb-mechanism-II} shows the time formation of
the six native H-bonds H1-H6 in the seven folding trajectories. 
As is shown in Fig.~\ref{fig:rmsd-extend-300k-s5} and
Fig.~\ref{fig:hb-mechanism-II} the two ends of
this $\beta$-hairpin come near to form H-bond H1 first (and to a lesser
extent H2 first), and the hydrophobic core reorganizes to its well packed state 
slowly. The partial helical structure
(Events 169 and 280) is not a necessary intermediate and appears in 4
of the 7 trajectories following mechanism II.  Comparing the two
mechanisms, one can see that they are not mutually
incompatible; many trajectories fall somewhere in between these two
descriptions. In some cases, for example, folding is initiated in the
middle region (H4). The $\beta$-sheet then propagates first outwards
(forming H1-H3) and then inwards (forming H5-H6) (see Run 6 in 
Fig.~\ref{fig:hb-mechanism-II}).

\begin{figure*}[ht!]

\vspace{-0.6cm}\includegraphics[width=13.0cm]{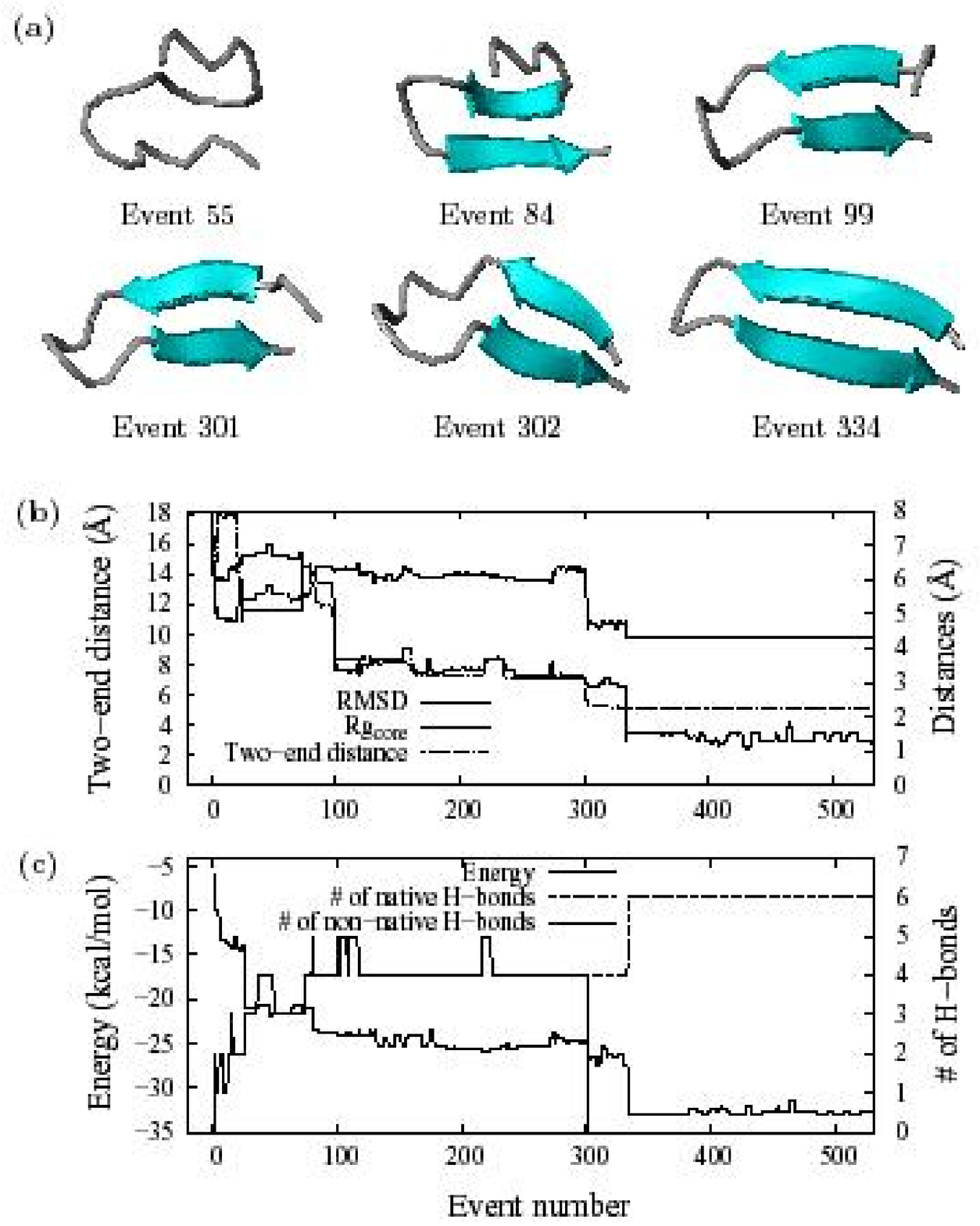}

\vspace{0.0cm}
\caption{
Detailed analysis of a folding trajectory following Mechanism III,
simulated at 300 K, starting from a fully extended state. Another 
folding trajectory was presented in a short communication.~\cite{WMD03}
(a) Six snapshots. 
(b) C$_\alpha$-rmsd from 2GB1 structure of 
the hairpin, radius of gyration of the hydrophobic core $Rg_{core}$, and the 
two-end distance as a function of accepted event number. 
(c) Total energy, number of all the native and non-native H-bonds in each 
sampled conformation as a function of accepted event number.
}
\label{fig:rmsd-extend-300k-s24}
\end{figure*}
\begin{figure}[ht!]
~\\ \vspace{-0.2cm}\includegraphics[width=8.2cm]{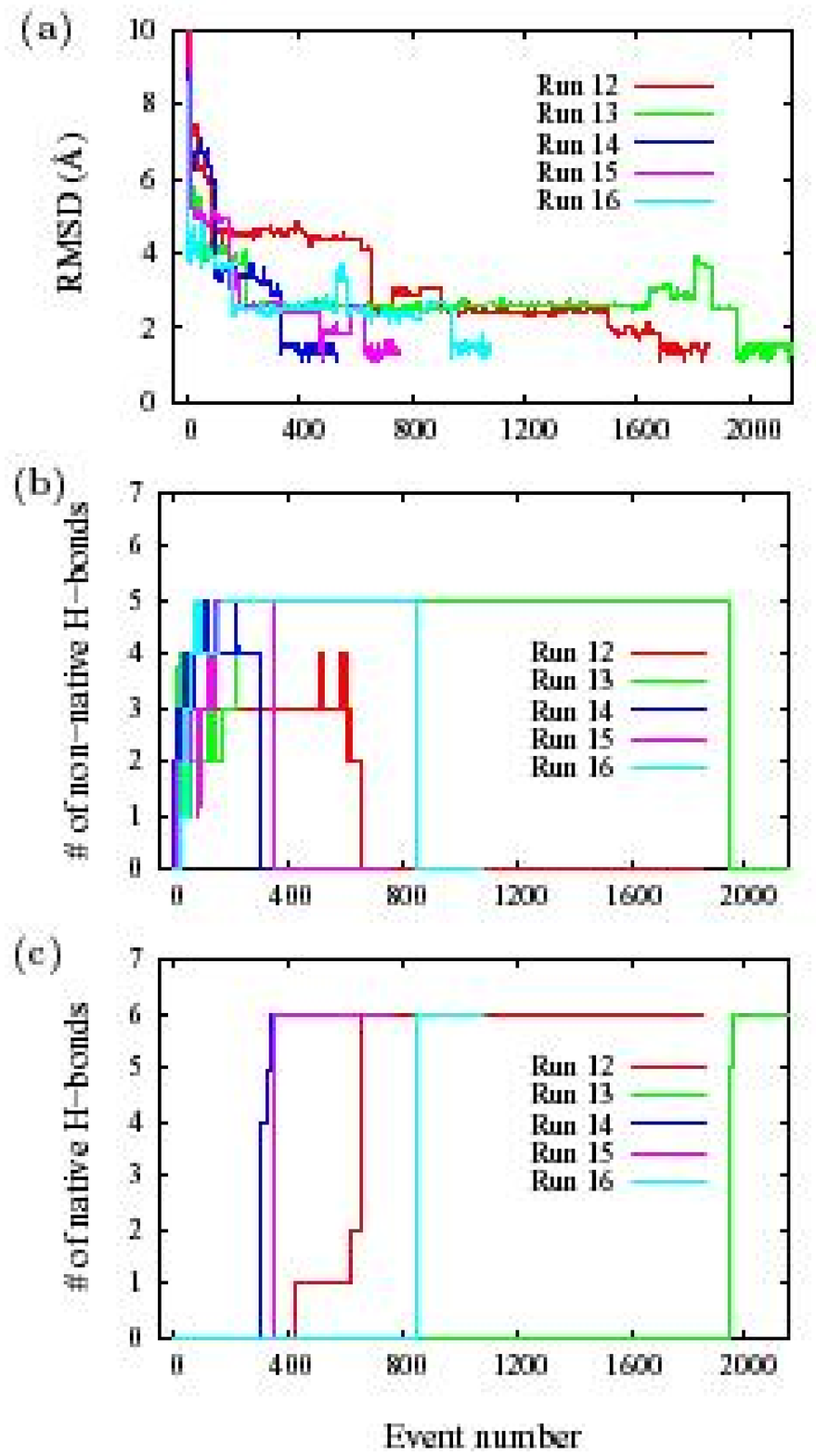}

\vspace{0.0cm}

\caption{
Analysis of all trajectories following Mechanism III. 
(a) Rmsd from the 2GB1 structure,  
(b) number of non-native H-bonds, and (c)  number of native H-bonds 
in each conformation as a function of accepted event number.  
Run 13 is the the folding simulation described in Fig. 3 in a short 
communication,~\cite{WMD03} and Run 14 is the folding simulation
described in Fig. 8. The final rmsd's from the 2GB1 structure are less
 than 1.8 \AA~(see (a)) in the five folding simulations.  In the five folding 
simulations, on average, the non-native H-bonds break almost 
at the same time (see(b)), then the native H-bonds form rapidly, almost 
instantaneously (see(c)).
}
\label{fig:rmsd-and-hb}
\end{figure}

Mechanism III  had not been observed
in previous all-atom folding,~\cite{PD01,ZHOU02} unfolding,~\cite{PD99} and
equilibrium simulations.~\cite{MK99,BS01,GS01} This mechanism, seen in
five folding trajectories, is characterized by a rapid folding into a
collapsed state with a turn at the wrong place, forming an asymmetric
hairpin structure stabilized by non-native H-bonds and a partially
packed hydrophobic core. Then slowly, step by step in a reptation
mode, the asymmetry is corrected, with non-native H-bonds breaking and
reforming, in a structure getting closer to the native hairpin. A
representative trajectory is given in Fig.~\ref{fig:rmsd-extend-300k-s24}.  
Folding begins with the
formation of a compact state defined by a partially packed
hydrophobic core and two non-native H-bonds 46:HN-55:O and 48:HN-53:O
(event 55). Then the number of non-native H-bonds increases to four:
46:HN-55:O, 48:HN-53:O, 53:HN-48:O, 55:HN-46:O, and a short
$\beta$-sheet structure (Event 84) appears. Driven by the hydrophobic
interactions, at event 99, the reptation motion of the loop causes the
four non-native H-bonds to break, and four new non-native H-bonds
44:HN-55:O, 46:HN-53:O, 53:HN-46:O, 55:HN-44:O form. This peptide
shows a new asymmetric $\beta$-sheet structure, which is closer to its
symmetric $\beta$-hairpin structure.  During the next 200 events, this
peptide stays in this state and reorganizes its hydrophobic
core. After an optimization of the hydrophobic and hydrogen bonding
interactions, at event 302, the reptation motion of the loop enhances
longitudinal motion of the two strands, breaking the four non-native
H-bonds and forming four native ones (H1-H4); simultaneously, the core
radius of gyration drops to 4.7~\AA. The hairpin structures forms
rapidly afterwards, with the addition of the fifth and sixth native
H-bond (H5 and H6), a rmsd dropping to 1.4~\AA\ (Event 334), and the
energy approaching to -33 kcal/mol.  Once the native hairpin forms, it
is fairly stable: the total number of H-bonds remains at six, the 
radius of gyration of the hydrophobic core keeps around 4.3~\AA, the 
rmsd fluctuates around 1.4~\AA, and the energy fluctuates around
$-$33 kcal/mol. As shown in Figure~\ref{fig:rmsd-extend-300k-s24} and
Fig.~\ref{fig:rmsd-and-hb}, the critical and rate-limiting step in this 
mechanism is the breaking, almost in synchrony, of four or five non-native 
H-bonds of a structure very close to the native state, followed by a rapid 
formation of the native H-bonds. It can also be seen from 
Fig.~\ref{fig:rmsd-extend-300k-s24} that the two ends of this $\beta$-hairpin 
slowly come near, and the hydrophobic core reorganizes to its well 
packed state slowly.

\section*{\bf DISCUSSION}

\subsection*{Sensitivity of Trajectories}

Because various interaction potentials have led previously to
conflicting reports regarding the details of folding, it is essential
to determine how changes in the force-field and the starting
structure affect the folding trajectories.

We consider here two variations of the force field. Firstly, we run 20
simulations using OPEP with an energy parameter for the H-bond
$\varepsilon_{hb}$ increased from the standard value of 1.5 to 2.5,
a change which could considerably affect the stability of intermediate
structures. Remarkably, we still recover the three folding mechanisms
among the six successful trajectories.  Secondly, we run 10
simulations using ART with G\={o}-like OPEP potential ( $E= E(\mbox{OPEP})+
k*(\mbox{rmsd}^2$), for $k=0.5$, 0.55 and 0.6) which favors native contacts in
the total energy. Starting from a fully extended state, all ten
simulations found the folded state following either mechanism I or II:
six simulations fold from the end region (H1 or H2 form first), two
from the middle (H3 forms first) and two from the turn (H5 first). As
in earlier G\={o}-model simulations,~\cite{ZHOU02} we find no helical
intermediates in any of the ten folding simulations.  
In these simulations, folding can be initiated at the end region, middle
region, and turn region of the peptide, in agreement with
the results obtained by using only the OPEP potential. Because native
interactions are strongly favored in G\={o}-model simulations, the
asymmetric conformations found in mechanism III become prohibitive,
preventing the appearance of mechanism III. The bias of the G\={o}
potential can therefore lead to an incomplete sampling of the folding
paths. 

\begin{figure*}[ht!]
\vspace{-0.6cm}\includegraphics[width=13.0cm]{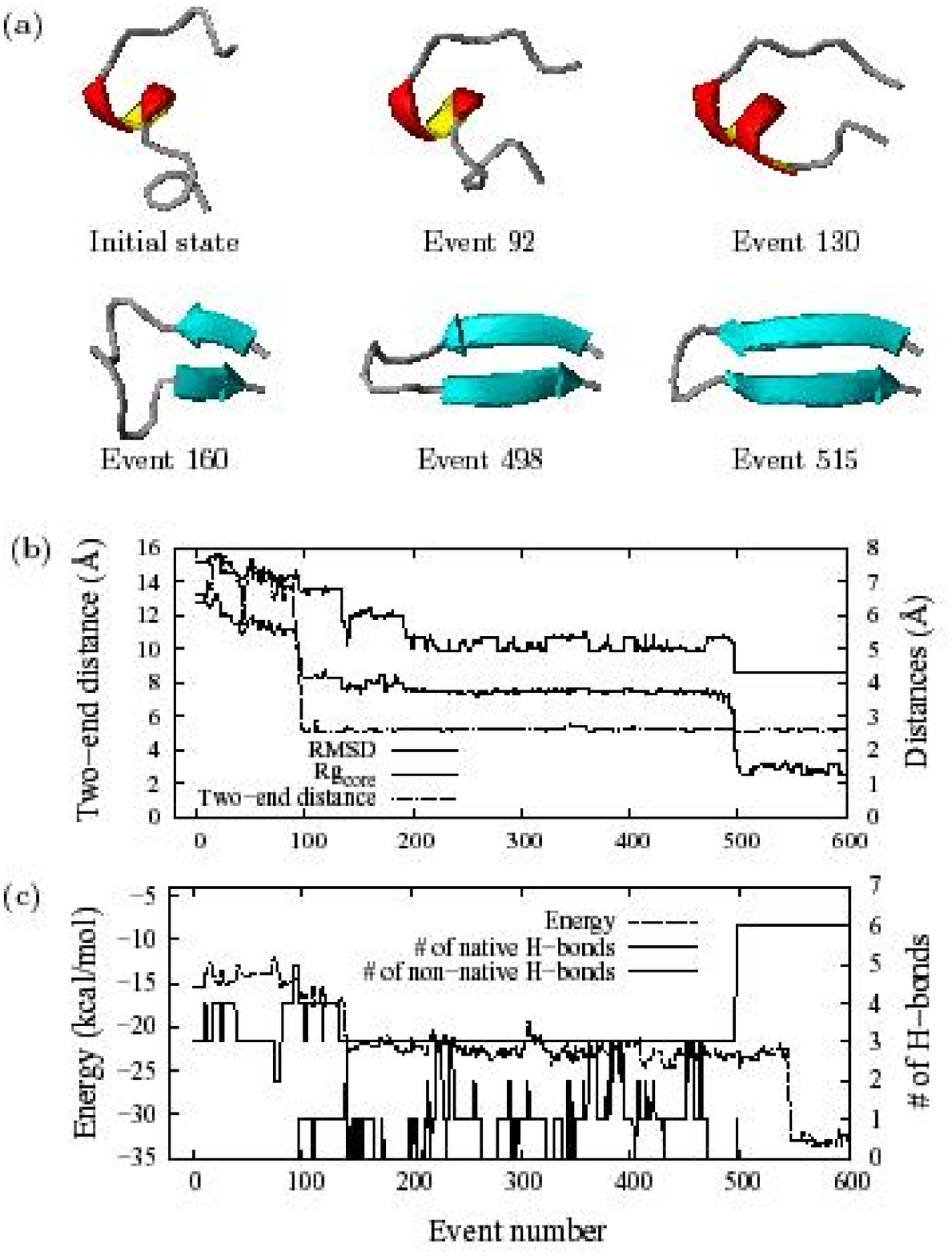}
\vspace{0.2cm}
\caption{
A detailed analysis of a trajectory resulting in the fully
$\beta$-hairpin at 300 K, starting from a semi-helical structure.
In this initial state, the 14 pairs of ($\phi$,$\psi$) values, excluding
residues Gly41 and Glu56, are 
(-63$^{\circ}$, 121$^{\circ}$), (-63$^{\circ}$, -27$^{\circ}$),
 (-65$^{\circ}$, 98$^{\circ}$), (-71$^{\circ}$, -10$^{\circ}$),  
(-48$^{\circ}$, 145$^{\circ}$),
 (-51$^{\circ}$, -61$^{\circ}$), (-56$^{\circ}$, -48$^{\circ}$),
 (-51$^{\circ}$, -55$^{\circ}$),
 (-61$^{\circ}$, -51$^{\circ}$), 
(-61$^{\circ}$, 94$^{\circ}$), (-63$^{\circ}$, -48$^{\circ}$),
 (-54$^{\circ}$, -43$^{\circ}$), (-59$^{\circ}$, -58$^{\circ}$),
 (-74$^{\circ}$, 80$^{\circ}$).
($\phi$,$\psi$) values near (-60$^{\circ}$, -45$^{\circ}$)
 are typical for $\alpha$-helices. 
(a) Six snapshots. (b) C$_\alpha$-rmsd from 2GB1 structure of 
the hairpin, radius of gyration of the hydrophobic core $Rg_{core}$, and the 
two-end distance as a function of accepted event number. 
(c) Total energy, number of all the native and non-native 
H-bonds in each sampled conformation as a function of accepted event number. 
}
\label{fig:rmsd-300k-s6}
\end{figure*}

To determine the impact of the starting conformation on the
ART-trajectories and to address the question of semi-helical
intermediates during folding,~\cite{PD01} eight trial simulations were
attempted at 300 K starting from a semi-helical structure.  Four of
those resulted in hairpin structures with six native H-bonds,
while the remaining four displayed helical structures or
$\beta$-hairpin structures with non-native H-bonds.
Interestingly, the four folded trajectories closely follow Mechanism
II.  Figure~\ref{fig:rmsd-300k-s6} gives a representation of folding
simulation at 300 K. The semi-helical structure rapidly relaxes to a
more compact structure (event 92) in which the two ends of the peptide
approach each other. At this stage, a partially packed hydrophobic
core ($Rg_{core}$ is 6.7 \AA) appears, without native H-bonds, however.  
Driven by the strong
hydrophobic interactions among the four hydrophobic residues (Tyr45,
Phe52, Trp43, Val54), the two ends of the peptide come nearer and one
native H-bond (H1) is formed at the end region of the peptide. After
50 events, the next two H-bonds H2 and H3 form almost
simultaneously. The peptide stays in this state for 350 events,
reorganizing slowly the hydrophobic core ($Rg_{core}$ oscillates around
5.2 \AA\ and the rmsd from the 2GB1 structure reaches a plateau (4
\AA)).  After a slow structural adjustment process, the hydrophobic
core forms fully in a large cooperative move at event 500 ($Rg_{core}$
drops to 4.3 \AA), and the six native H-bonds set in rapidly
afterwards (H4 and H5 form first, followed by the formation of
H6). This event marks the completion of the folding process of the
peptide with the rmsd reaching to 1.5 \AA.  This demonstrates that
ART can find the folded state starting from a helical structure and
that this helical structure may exist in the real folding pathway of this 
protein, as discussed in previous simulations~\cite{GS01,PD01,IRB03} and recent
energy landscape characterization of $\beta$-hairpin2 and its
isomers~\cite{NUS03}.

From the structure of the initial helical state, we can explain why the four
folded trajectories only follow Mechanism II.  This state is characterized
 by  
a helical structure spanning residues 47-50, and two non-native H-bonds
(50:HN-46:O and 51:HN-47:O). The existence of
this helical segment makes it difficult to form native H-bonds at
the turn or the middle region of the peptide because of geometric
restrictions. However the two ends of this peptide are very flexible,
driven by the hydrophobic interactions, they can approach each other
easily, then form a H-bond first between them. When the initial state
is a fully extended state, this peptide has much more freedom to find
its native state, and this is why multiple folding pathways are
present.
 
\subsection*{Cluster Analysis of the Folded Trajectories}
 
As we have seen in the previous section, the folding process can be
described by a single concept: the competition between three types of
interactions. A cluster analysis shows, however, that this does not
mean that the folding trajectories can be unified. 

We perform a cluster analysis following the procedure described in
Ref.~\onlinecite{Daura} (see Methods), using a C$_{\alpha}$-rmsd 
cutoff of 1.5 \AA.
All the accepted conformations in each folded trajectory of the 
26 folding simulations obtained by the standard OPEP potential are used 
for this procedure. A total of 12-40 clusters were found for each folded 
trajectory, indicating strong variations in the details of the folding trajectories.
Moreover, we find very little overlap of the basins between similar trajectories: 
except for the trivial initial and native clusters, it is generally 
not possible to match more than one or two clusters between trajectories 
following the same mechanisms. We obtained the same qualitative results 
using other clustering analysis. 

The failure of cluster analysis  by C$_{\alpha}$-rmsd  to characterize the three 
different folding mechanisms for this  $\beta$-hairpin is due to the flexibility of 
this small peptide. For example, the asymmetric $\beta$-hairpin structures  in 
the folding trajectory following the reptation mechanism are very diverse, 
presenting a wide range of hydrogen-bond patterns. Moreover, the 
C$_{\alpha}$-rmsd between a beta hairpin with the turn shifted to C-terminal 
and a beta hairpin with the turn shifted to N-terminal is as big as 4.3 \AA. 
Increasing the C$_{\alpha}$-rmsd cutoff will make it difficult to differentiate 
asymmetric $\beta$-hairpin state from the folded state.  The classification in terms 
of folding mechanisms as identified by the formation of hydrogen bonds appears 
therefore superior to the clusterization method for this small peptide.

\subsection*{Does the Reptation Folding Trajectory Exist?}

Surprizingly, although present in part in many previous simulations,
the reptation mechanism had not been identified previously.  The
asymmetric conformations with only non-native H-bonds, which
characterize the intermediate states in Mechanisms III, are found in
many simulations.  For example, a cluster analysis of the structures
produced in an all-atom multicanonical MC simulation of the $\beta$-hairpin2
finds that these asymmetric conformations account for 20\% of all
conformations.~\cite{MK99} Similarly, Skolnick and collaborators find
that a lattice model often folds into asymmetric structures.~\cite{KI99}
In a recent work of Irb\"ack, a local minimum corresponding to a $\beta$-hairpin 
with non-native topology is observed in the energy landscape of this 
$\beta$-hairpin2.~\cite{IRB03}
Finally, most of the folded conformations identified by Zagrovic {\it
et al.}  in distributed MD simulations (Folding@home) seem
asymmetric although this is not explicitly stated.~\cite{PD01} 
Similar results are found in smaller peptides such
as the 11-residue model peptide of Wang {\it et al.} studied by
molecular dynamics, which also displays asymmetric $\beta$-hairpin
structures.~\cite{WJ99} 

Although many simulations found the asymmetric conformation, none
seems to have been able to overcome the rate-limiting step, which requires
breaking all non-native bonds at once (Fig.~\ref{fig:rmsd-and-hb} (b)), in 
order to form the native state. With OPEP, this barrier is found to be about 
12 kcal/mol and corresponds to a time scale on the order of 
$\mu$s,~\cite{DS98} near the experimental folding time but much beyond 
what can be reached by standard molecular dynamics. 

In addition to display a time scale in agreement with experiment,
there is experimental evidence that many $\beta$-hairpin sequences can
populate two distinct hairpin conformations of various loop lengths
and pairings of $\beta$-strands in solution.~\cite{SE95,RA99}
Fluorescence microscopy also suggests that myosin can induce the
reptation of actin filaments when adenosine triphosphate is
added.~\cite{HD02} The reptation mechanism, which has been well
established as a fundamental movement in polymer chain, might
therefore exist in the folding of a simple $\beta$-hairpin.

\subsection*{Competition between Hydrophobic and Native Hydrogen 
Bonding  Interactions}

The folding mechanisms described above are the result of a strong
competition between hydrophobic and native hydrogen-bonding
interactions.  All cases show that the hydrophobic
interactions play a dominant role in the folding process. At the
beginning of the folding, a partially packed hydrophobic core always
forms before native H-bonds appear (see
Fig.~\ref{fig:rmsd-extend-300k-s4}(b) and (c),
Fig.~\ref{fig:rmsd-extend-300k-s5}(b) and (c),
Fig.~\ref{fig:rmsd-extend-300k-s24}(b) and (c),
Fig.~\ref{fig:rmsd-300k-s6}(b) and (c)). The following step is the
rearrangement of the hydrophobic core and the optimization between the
complete hydrophobic and native hydrogen bonding interactions. When
hydrophobic and hydrogen bonding interactions reach a balance, the
well packed hydrophobic core forms before
(Fig.~\ref{fig:rmsd-extend-300k-s4}(b) and (c)) or at the same time as
(Fig.~\ref{fig:rmsd-extend-300k-s5}(b) and (c), 
Fig.~\ref{fig:rmsd-extend-300k-s24}(b) and (c),
Fig.~\ref{fig:rmsd-300k-s6}(b) and (c)) the native H-bonds network forms.

\subsection*{Competition between Native and Non-native Hydrogen 
Bonding Interactions}

Native and non-native hydrogen-bonding interactions also compete strongly
during the folding process.  The initial collapse is always
accompanied by the formation of three to six non-native H-bonds (see
Fig.~\ref{fig:rmsd-extend-300k-s4}(c),
Fig.~\ref{fig:rmsd-extend-300k-s5}(c),
Fig.~\ref{fig:rmsd-and-hb}(b),
Fig.~\ref{fig:rmsd-300k-s6}(c)). However, these non-native H-bonds
are not stable in the long run; they form, break, and reform in
response to the movement of the hydrophobic core.  Driven by the
hydrophobic interactions, the non-native hydrogen bonding interactions
finally become weaker, and native H-bonds form, leading rapidly to
the native state. The whole folding process can therefore be described 
as a balance between hydrophobic, native and non-native hydrogen 
bonding forces.

\section*{\bf CONCLUSIONS}

By demonstrating that the folding of a $\beta$-hairpin can be
initiated at the end, the middle and the turn region, as well as from
an asymmetric conformation, the three folding mechanisms proposed here
help reconcile conflicting theoretical data on the hairpin2 of protein
G~\cite{EATON,MK99,THI00,ZHOU02} or between various hairpins, e.g. the
first hairpin of tendamistat.~\cite{BON00}
Using these three mechanisms, we can now propose a complete picture
of the folding of $\beta$-hairpins which does not depend on the exact
amino-acid composition. The exact folding path followed by a given
$\beta$-hairpin should be influenced by its sequence and the solvent
conditions; all paths should, however, belong to one of the three
mechanisms presented here. The first two mechanisms, with the
propagation from either the turn or the end points, had already been
described in previous reports on $\beta$-hairpin 2, but no previous method 
had managed to detect both pathways. The third mechanism, identified in
these simulations for the first time, involves folding into an
asymmetric state followed by a reptation of one strand over the other
until the peptide reaches its native state.  The existence of this
mechanism is suppported by a number of experimental and numerical
results even though the rate-limiting step and the presence of
non-native states place it outside the scope of most simulation
methods, including unfolding, G\={o} and standard MD approaches. This
last results underlines the importance of direct non-biases methods,
such as ART, for studying the folding process.

Although complex, presenting a large number of different paths, the
folding of this $\beta$-hairpin can still be described by a unique
process of competition between hydrophobic core and native and
non-native H-bond interactions.  In particular, it is clear that
non-native H-bond interactions can play a critical role in the
folding process even though they are absent in the final product. 

\section*{\bf ACKNOWLEDGEMENTS}

GW and NM are supported in part by the {\it Fonds qu\'eb\'ecois pour
la formation des chercheurs et l'aide \`a la recherche} and the {\it
Natural Sciences and Engineering Research Council} of Canada. Most
of the calculations were done on the computers of the {\it R\'eseau
qu\'eb\'ecois de calcul de haute performance} (RQCHP). NM is a
Cottrell Scholar of the Research Corporation. We thank Drs. Hue Sun
Chan, Marek Cieplak, and Saraswathi Vishveshwara for useful discussion.


\begin{thebibliography}{10}

\bibitem{NMR95}
Blanco FJ, and Serrano L.
\newblock Folding of protein GB1 domain studied by the conformational
  characterization of fragments comprising its secondary structure elements.
\newblock {Eur J Biochem} 1999;{230:}634--649. 

\bibitem{EATON} 
Munoz V, Thompson PA, Hofrichter J, and Eaton WA. 
\newblock Folding dynamics and mechanism of beta-hairpin formation.
\newblock {Nature} 1997;{390:}196--199.

\bibitem{DU98}
Duan Y, and Kollman PA. 
\newblock Pathways to a protein folding intermediate observed in a
  1-microsecond simulation in aqueous solution.
\newblock {Science} 1998;{282:}740--744.

\bibitem{MUN98}
Munoz V, Henry ER, Hofrichter J, and Eaton WA. 
\newblock A statistical mechanical model for beta-hairpin kinetics.
\newblock {Proc Natl Acad Sci USA} 1998;{95:}5872--5879.

\bibitem{KI99}
Kolinski A, Ilkowski B, and Skolnick J.
\newblock Dynamics and thermodynamics of beta-hairpin assembly: insights from
various simulation techniques.
\newblock {Biophys J} 1999;{77:}2942--2952.

\bibitem{THI00}
Klimov DK, and Thirumalai D. 
\newblock Mechanisms and kinetics of $\beta$-hairpin formation.
\newblock {Proc Natl Acad Sci USA} 2000;{97:}2544--2549.

\bibitem{TSAI02}
Tsai J, and Levitt M. 
\newblock Evidence of turn and salt bridge contributions to $\beta$-hairpin
  stability: MD simulations of C-terminal fragment from the B1 domain of
  protein G.
\newblock {Biophys Chem} 2002;{101-102:}187--201.

\bibitem{PD01}
Zagrobic B, Sorin EJ, and Pande V.
\newblock $\beta$-hairpin folding simulations in atomistic detail using an
  implicit solvent model.
\newblock {J Mol Biol} 2001;{313:}151--169.

\bibitem{BS01}
Zhou R, Berne BJ, and Germain R.
\newblock The free energy landscape for $\beta$ hairpin folding in explicit
  water.
\newblock {Proc Natl Acad Sci USA} 2001;{98:}14931--14936.

\bibitem{ZHOU02}
Zhou Y, and Linhananta A. 
\newblock Role of hydrophilic and hydrophobic contacts in folding of the second
  $\beta$-hairpin fragment of protein G: molecular dynamics simulation studies
  of an all-atom model.
\newblock {Proteins} 2002;{47:}154--162.

\bibitem{PD99}
Pande VS, and Rokhsar DS. 
\newblock Molecular dynamics simulations of unfolding and refolding of a
  beta-hairpin fragment of protein G.
\newblock {Proc Natl Acad Sci USA} 1999;{96:}9062--9067.

\bibitem{LEE01}
Lee J, and Shin S. 
\newblock Understanding $\beta$-hairpin formation by molecular dynamics
  simulations of unfolding.
\newblock {Biophys J} 2001;{81:}2507--2516.

\bibitem{MK99}
Dinner AR, Lazaridis T, and Karplus M. 
\newblock Understanding $\beta$-hairpin formation.
\newblock {Proc Natl Acad Sci USA} 1999;{96:}9068--9073.

\bibitem{GS01}
Garc\'{i}a AE, and Sanbonmatsu KY. 
\newblock Exploring the energy landscape of a $\beta$-hairpin in explicit
  solvent.
\newblock {Proteins} 2001;{42:}345--354.

\bibitem{MA00}
Ma B, and Nussinov R. 
\newblock Molecular dynamics simulations of a beta-hairpin fragment of protein
  G: balance between side-chain and backbone forces.
\newblock {J Mol Biol} 2000;{296:}1091--1104.

\bibitem{IRB03}
Irb\"ack A, Samuelsson B, Sjunnesson F, and Wallin S. 
\newblock Thermodynamics of $\alpha$- and $\beta$-structure formation in proteins.
\newblock {Biophys J} 2003;{85:}1466--1473.

\bibitem{AF02}
Fersht AR. 
\newblock On the simulation of protein folding by short time scale molecular
  dynamics and distributed computing.
\newblock {Proc Natl Acad Sci USA} 2002;{99:}14122--14125.

\bibitem{BM96}
Barkema GT, and Mousseau N. 
\newblock Event-based relaxation of continuous disordered systems.
\newblock {Phys Rev Lett} 1996;{77:}4358--4361.

\bibitem{BM98}
Barkema GT, and Mousseau N. 
\newblock Identification of relaxation and diffusion mechanisms in amorphous
  silicon.
\newblock {Phys Rev Lett} 1998;{81:}1865--1868.

\bibitem{FOR01}
Forcellino F, and Derreumaux P. 
\newblock Computer simulations aimed at structure prediction of supersecondary
  motifs in proteins.
\newblock {Proteins} 2001;{45:}159--166.

\bibitem{WMD02}
Wei G, Mousseau N, and Derreumaux P.
\newblock Exploring the energy landscape of proteins: A characterization of the
  activation-relaxation technique.
\newblock {J Chem Phys} 2002;{117:}11379--11387.

\bibitem{BON00}
Bonvin AM, and van Gunsteren WF. 
\newblock $\beta$-hairpin stability and folding: molecular dynamics studies of
  the first $\beta$-hairpin of tendamistat.
\newblock {J Mol Biol} 2000;{296:}255--268.

\bibitem{WJ99}
Wang H, Varady J, Ng L, and Sung S. 
\newblock Molecular dynamics simulations of $\beta$-hairpin folding.
\newblock {Proteins} 1999;{37:}325--333.

\bibitem{WMD03}
Wei G, Derreumaux P, and Mousseau N.
\newblock Sampling the complex energy landscape of a simple $\beta$-hairpin.
\newblock {J Chem Phys} 2003;{119:}6403-6406.

\bibitem{MM00}
Malek R, and Mousseau N. 
\newblock Dynamics of Lennard-Jones clusters: A characterization of the
  activation-relaxation technique.
\newblock {Phys Rev E} 2000;{62:}7723--7728.

\bibitem{MDB01}
Mousseau N, Derreumaux P, Barkema GT, and Malek R. (2001).
\newblock Sampling activated mechanisms in proteins with the
  activation-relaxation technique.
\newblock {J Mol Graph Model} 2001;{19:}78--86.

\bibitem{Lan88}
Lancz\'os C. 
\newblock {Applied Analysis}.
\newblock Dover, New York; 1988.

\bibitem{Wales99}
Munro LJ, and Wales DJ.
\newblock Defect migration in crystalline silicon.
\newblock {Phys Rev B} 1999;{59:}3969--3980.

\bibitem{DP99}
Derreumaux P. 
\newblock From polypeptide sequences to structures using Monte Carlo
  simulations and an optimized potential.
\newblock {J Chem Phys} 1999;{11:}2301--2310.

\bibitem{DP00}
Derreumaux P. 
\newblock Generating ensemble averages for small proteins from extended
  conformations by monte carlo simulations.
\newblock {Phys Rev Lett} 2000;{85:}206--209.

\bibitem{GF91}
Gronenborn AM, and et~al. 
\newblock A novel, highly stable fold of the immunoglobulin in binding domain
  of streptococcal protein-G.
\newblock {Science} 1991;{253:}657--661.

\bibitem{KS83}
Kabsch W, and Sander C. 
\newblock Dictionary of protein secondary structure: pattern recognition of
  H-bond and geometrical features.
\newblock {Biopolymers} 1983;{22:}2577--2637.

\bibitem{MOL96}
Koradi R, Billeter M, and Wuthrich K. 
\newblock Molmol: A program for display and analysis of macromolecular
structures.
\newblock {J Mol Graphics} 1996;{14:}51--55.

\bibitem{NK02}
Kamiya N, Higo J, and Nakamura H. 
\newblock Conformational transition states of $\beta$-hairpin peptide between
  the ordered and disordered conformations in explicit water.
\newblock {Protein Sci} 2002;{11:}2297--2307.

\bibitem{NUS03}
Ma B, and Nussinov R. 
\newblock Energy landscape and dynamics of the $\beta$-hairpin G peptide and 
  its isomers: Topology and sequences.
\newblock {Protein Sci} 2003;{12:} 1882--1893.

\bibitem{Daura}
Daura X, van Gunsteren WF, and Mark AE.
\newblock Folding-unfolding thermodynamics of a heptapeptide from equilibrium
  simulations.
\newblock {Proteins} 1999;{34:}269--280.

\bibitem{DS98}
Derreumaux P, and Schlick T. 
\newblock The loop opening/closing motion of the enzyme triosephosphate
  isomerase.
\newblock {Biophys J} 1998;{74:}72--81.

\bibitem{SE95}
Searle MS, Williams DH, and Packman LC. 
\newblock A short linear peptide derived from the N-terminal sequence of
  ubiquitin folds into a water-stable non-native beta-hairpin.
\newblock {Nat Struct Biol} 1995;{2:}999--1006.

\bibitem{RA99}
Ramirez-Alvarado M, Kortemme T, Blanco FJ, and Serrano L. (1999).
\newblock beta-hairpin and beta-sheet formation in designed linear peptides.
\newblock {Bioorganic Medicinal Chemistry}{ \bf 7}, 93--103.

\bibitem{HD02}
Humphrey D, Duggan C, Saha D, Smith D, and K\"{a}s J. 
\newblock Active fluidization of polymer networks through molecular motors.
\newblock {Nature} 2002;{416:}413--416.

\end{thebibliography}
\end{document}